\documentclass[12pt,a4paper]{article}
\usepackage[dvips]{epsfig}
\usepackage[latin1]{inputenc}
\usepackage[english]{babel}
\usepackage{graphicx} 
\DeclareGraphicsExtensions{.eps,.ps,.eps.gz,.ps.gz,.eps.Z}

\begin{document}

\thispagestyle{empty}

\title{Reconstruction of the Free Energy in the Metastable Region using the Path Ensemble}

\author{Armando Ticona Bustillos$^*$, Dieter W. Heermann \\and \\
     Claudette E. Cordeiro\footnote{Permanent address: Instituto de Fisica, Universidade Federal Fluminense, 24.210.340-Niteroi-RJ-BRAZIL}
         \\ {} \\
           Institut f\"ur Theoretische Physik \\
           Universit\"at Heidelberg \\
           Philosophenweg 19 \\
           D-69120 Heidelberg\\
           and\\ 
           Interdisziplin\"ares Zentrum\\
           f\"ur Wissenschaftliches Rechnen\\
           der Universit\"at Heidelberg \\ {} \\
\vspace {3ex}}
\newpage

\maketitle
\vfill\eject
\begin{abstract}
By quenching into the metastable region of the three-dimensional
Ising model,  we investigate the paths that the magnetization
(energy) takes as a function of time. We accumulate  the
magnetization (energy) paths into time-dependent distributions
from which we reconstruct the  free energy as a function of  the
magnetic field, temperature and system size. From the
reconstructed free energy, we  obtain the free energy barrier that
is associated with the transition from a metastable  state to the
stable equilibrium state. Although mean-field theory predicts a
sharp transition  between the metastable and the unstable region
where the free energy barrier is zero, the  results for the
nearest-neighbour Ising model show that the free energy barrier
does not  go zero.
\end{abstract}

\vspace{2cm} {\bf Keywords:} first-order phase transition, nucleation, spinodal, free energy, free energy barrier
non-equilibrium statistical physics, Monte Carlo simulation, kinetic Ising model, scaling, 
mean-field theory, master equation

\vfill\eject

\section{Introduction}
\boldmath

Of considerable interest is the calculation of the free energy in the meta\-stable
region of a system with a first-order phase transition. The free energy is the
starting point for many theories dealing with metastability or spinodal
decomposition. The free energy barrier height is needed to calculate the cost 
of creating a critical droplet
formed by  statistical fluctuations, which starts the
transformation of a metastable state to a stable equilibrium state
[1-5]. The calculation of the free energy is therefore  central in
the development of an understanding of metastability, the
spinodal, and the phenomena that arise during a first-order phase
transition. It is therefore unfortunate that the free energy
cannot be directly measured in a computer
simulation~\cite{Bennett}.

\begin{figure}[ht]
\begin{center}
{\epsfig{file=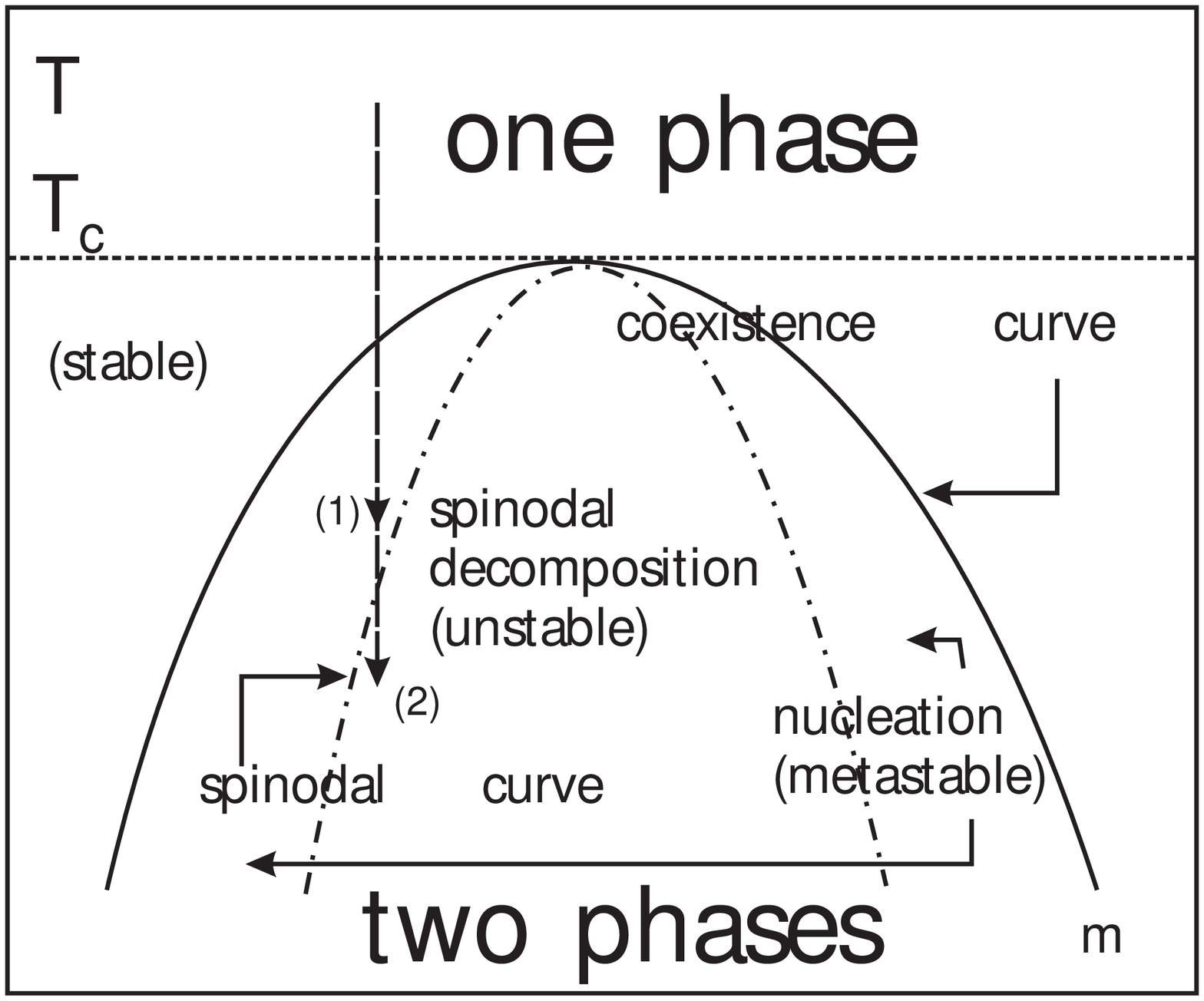,width=12cm}}
\caption{\label{fig:ab-phase} Schematic ($T-m$) phase diagram.
Indicated is a temperature quench. Classical theories suggest that
relaxation towards equilibrium following a quench in the indicated region (1) proceeds
by nucleation and growth. From mean-field theories for the free
energy one also obtains a spinodal which separates the metastable
region from the unstable}
\end{center}
\end{figure}

In mean-field theories [5-8] (infinite-range interaction)
for the first-order phase transition,
one can calculate the free energy not only for the stable
equilibrium but also for the non-equilibrium case. It is from this
calculation that the distinction is made between metastable and
unstable states (see Figure~\ref{fig:ab-phase}). The meta- and 
unstable states are separated in this
framework by a spinodal where the free energy barrier is zero. 
Within the spinodal region the homogeneous (disordered) phase is
thermodynamically unstable.  Between the spinodal and the coexistence curve one
needs nucleation events of the opposite phase  (thermal
activation, droplets) to induce the phase transformation.

It has been shown that for systems with short-range interactions,
fluctuations will always lead  to the decay of a metastable state
and drive the system to thermal equilibrium~\cite{Sewell,Heermann-Klein}. 
This does, however, not answer the question of the existence of a
spinodal. There have been several attempts to obtain a free energy
covering equilibrium and non-equilibrium. Notably Kaski et
al.~\cite{Kaski} tried to construct the free energy using a
coarse-graining procedure. A similar line of reasoning using
static concepts is the analytic continuation~\cite{Langer} of the
equilibrium free energy into the two-phase region.

Another approach to the problem of nucleation and metastability
has been taken by Binder~\cite{Binder-drop}. He
considered a single droplet in a finite volume and  analysed the
corresponding first order phase transition. This, however, does not
address the issue whether a spinodal exists and is not able to
reconstruct the free energy. It rather  assumes a geometric model
for the fluctuation and allows comparison to classical theories
of nucleation~\cite{Becker,Zettlemover}. Here we do not want to make 
any geometric assumptions for the nucleation events or the functional 
form of the free energy. 

Kolesik et al.~\cite{Mark} can also not reconstruct the spinodal
but offer an interesting approach to obtaining the lifetime of
metastable states. They look at what they call the projective
dynamics~\cite{Mark2}.

For a quantity such as the energy it is sufficient to take an
average over a small representative sample of states, but for the
free energy it is necessary to  consider all the states accessible
to the system. In this work we use the relaxation paths  of the
system to obtain functionals of the trajectory that the system
takes through  phase-space~\cite{Paul1, Paul2}. Hence we explore
the states accessible to the system escaping a metastable state
after a quench. We
use the path function to reconstruct the  free energy, using the
fact that an average of a path function is implicitly an average
over a suitably defined ensemble of paths.

We thus start with a description of how we obtain the paths in our
Monte Carlo simulation of the three-dimensional Ising model. We
use this model because a large body of data is available  to
facilitate comparision. We discuss the choices  for the transition
probability which will influence the dynamics of the system and
thus the escape from non-equilibrium. We reconstruct the free
energy from our simulation data, compare the computed
barrier height to mean-field theory and discuss the implications.

\section{The Free Energy calculated by paths}

The model that we use to study the first-order phase transition is
the Ising model. The  Hamiltonian of the Ising model for a simple 
cubic lattice $L^3=N$ is defined by 
\begin{equation}
{\cal H}(s) = - J \sum_{<ij>}s_is_j\quad - \mu H\sum_{i}s_i, \qquad
s_{i} = \pm 1
\end{equation}

\noindent where $J>0$ ($\frac{J}{k_B T_c} = 0.221673$~\cite{kt}) is the exchange
coupling, 
$h=2\frac {\mu H}{kT}$ is a dimensionless magnetic field, and
$s=(s_1, s_2, \ldots , s_i, \ldots ,s_N)$  denotes a configuration
of spins for the lattice. We studied the temperatures $T/T_c = 0.55$, 0.59, and
$0.63$ and fields $h$ ranging from $h=0.4$ to $0.85$.

The above defined Ising model does not have any intrinsic
dynamics. It is therefore necessary to construct a Monte Carlo
dynamics for the system. The dynamics of the system with respect 
to Monte Carlo simulations is specified by the transition
probabilities of a Markov chain  that establishes a path through
the available phase space. We have used the Metropolis transition 
probability~\cite{Heermann1,Binder-Heermann,Glauber}

\begin{equation}
W_M[s_i,E_i] = \frac{1}{\tau}\min \{1,\exp(-2\beta s_i
E_j)\}= W_M(s_i|s),
\end{equation}

\noindent where $\beta = 1/k_B T$ and the Glauber transition
probability~\cite{Heermann1, Binder-Heermann,Glauber}

\begin{equation}
W_G[s_i,E_i] = \frac{1}{\tau}(1-\tanh (\beta s_iE_j) =
W_M(s_i|s).
\end{equation}

$\tau$ is a constant setting the time scale. $E_j$ denotes the spin and the local field
before the spin flip. Here we have excluded the probability for
the suggestion of a new state which in our case is a constant. For
equilibrium properties it is sufficient that the dynamics satisfies
detailed balance so that the correct equilibrium distribution is
generated~\cite{Deem}.  Starting from an initial configuration
$s_0$, we get a sequence

\begin{equation}
            s^k_0,s^k_1,\ldots s^k_{n-1}
\end{equation}

\noindent of $n$ samples of spin configurations which are dynamically correlated. $k$ denotes the
sample of the path. We always start from the same initial condition (all spins down). However, 
since the random numbers change from run to run, we get a new path for each sample.

Time $t$ in this context is measured in Monte Carlo steps per spin. One Monte Carlo step (MCS) per 
lattice site, that is, one sweep through the entire lattice,
comprises one time unit. Neither magnetization  nor energy is
conserved in the model, which makes possible to compute energy $e$  and magnetization
$m=(1/N)\sum_{i=1}^Ns_i$ as a function of temperature, applied
field, time, as well as  the system size $N$.

From the point of view of the transition probabilities the dynamic
interpretation of the Monte Carlo  simulation algorithm stems from
the master equation

\begin{equation}
\frac{\partial P(s,t)}{\partial t} = \sum^N_i W(s|s_i)P(s_i,t) -
P(s,t)\sum^N_i W(s_i|s),
\end{equation}

\noindent where the condition of final equilibrium is guaranteed by the above choices of the
transition probabilities.

\begin{figure}[ht]
\begin{center}
{\epsfig{file=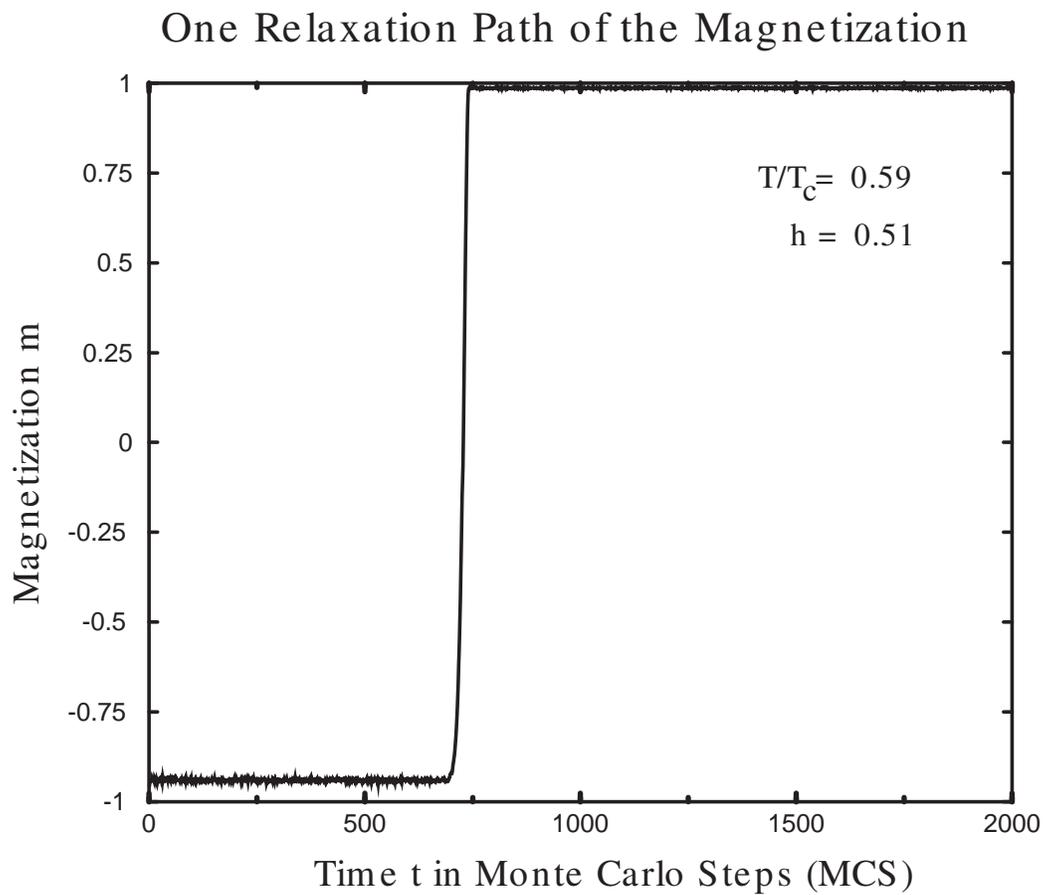,width=14cm}}
\caption{\label{fig:relaxation_path} Shown is one example of a relaxation path which is
used for the compilation of the probability distribution $P_N(m,e,t)$ }
\end{center}
\end{figure}

We now start with a configuration of spins where all spins are $-1$ ($m=-1$) as the
initial condition. This corresponds to an equilibrium state of the system. From this
state we quench, using the magnetic field $h$ opposite to the magnetization, into the 
two phase region, that is, to a non-equilibrium  magnetization 
$-m_{coex}(T)\le m_{meta}\le m_{coex}(T)$. 
If $h$ is small enough, the system settles
into a metastable state. In Figure~\ref{fig:relaxation_path} we
show such a path the magnetization takes from the stable into
metastable and then equilibrium state.

Following the magnetization values in time $t$, we get 

\begin{equation}
m(t) = (1/N)\sum_i s_i 
\end{equation}

\noindent for a single quench. For a single quench the values fluctuate after
an initial time-lag around a metastable quasi-equilibrium value. After
time the magnetization sharply changes its sign and settles into the
stable equilibrium value. 

The same behaviour is seen for the dimensionless energy per volume 
as a function of time

\begin{equation}
e(t) = (1/N){\cal H}(s(t)) .
\end{equation}

Repeating the quench with different initial conditions (here
different values for the seed of the random number generator) we
obtain a sample of all possible paths from the starting
equilibrium state, via the metastable state to the final
equilibrium state.

\begin{figure}[ht]
\begin{center}
{\epsfig{file=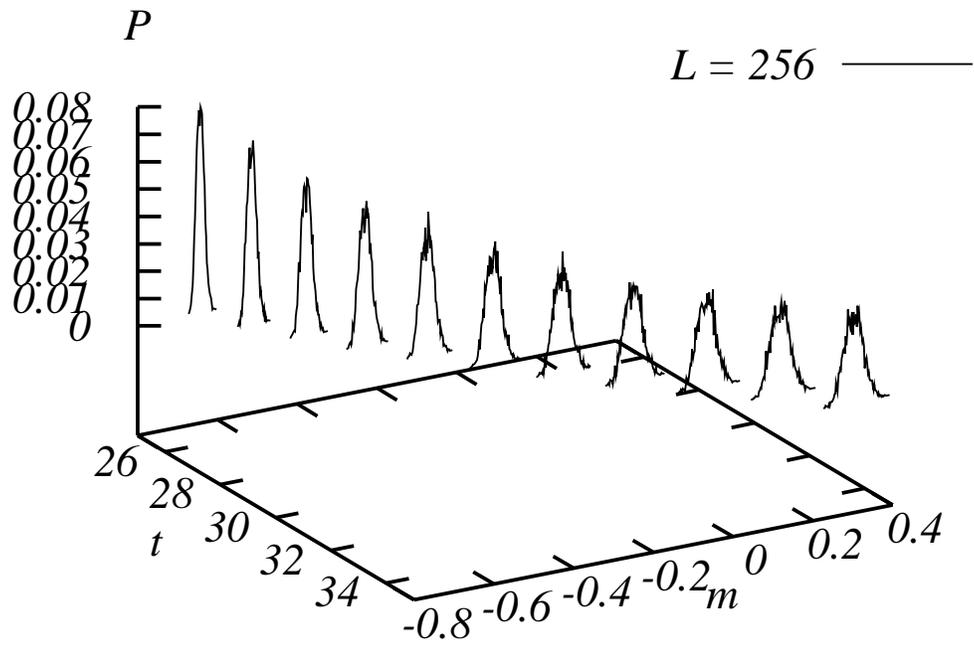,width=14cm}}
\caption{\label{fig:dist} Evolution of the distribution of the magnetization at $T/T_c=0.59$}
\end{center}
\end{figure}

For all the quenches we thus get the time-dependent distribution $P(m,e,N)(t)$ of
the magnetization and energy.
In Figure~\ref{fig:dist} we show the evolution of  $P(m,N)(t)$
(obtained by ignoring the energy spectrum) for the linear system 
sizes $L=32$, 64, 128, 265. For $L=32$ the distribution of the
magnetization values is rather broad and sharpens considerably
with increasing system size. After the initial relaxation the
distribution of $P(m,N)(t)$ shows a pronounced peak indicating
that the system has spent considerable time in a metastable
state. As time progresses, we pick up more contributions from
those paths that have already left the average magnetization of
the metastable state and are en route to equilibrium.  

In equilibrium one can obtain the free energy $F$ of this system
by computing the partition function 

\begin{equation}
Z = \sum_s \exp (-{\cal H}(s)/k_BT) 
\end{equation}

\noindent to obtain $F$ 

\begin{equation}
F = -k_BT \ln (Z) .
\end{equation}

\noindent The sum is extended over all possible $2^N$ spin configurations.

Here we use the time-dependent distributions for the magnetization
and the energy  to compute the free energy. We look at the
non-equilibrium situation that results after a  quench from a
stable equilibrium state to a state in the two phase region. The
system exhibits a quasi-stable behaviour before it finally reaches
stable equilibrium again.  We now make use of the dynamics of
the system. Starting from equilibrium we perturb the system,
applying a magnetic field $h$ opposite to the magnetization $m$,
and follow the path the system takes back to equilibrium.
Following many such paths we get the path ensemble as defined by
the initial thermal equilibrium and the process by which the
system is subsequently perturbed from that equilibrium.
Distributions and averages are then taken over the ensemble of
paths generated by this process.

We define the partition function for each value of the
magnetization resulting from the distribution of energy and
magnetization by summing over all energy values and over all $t$
in much the same way as  was done by~\cite{path} 

\begin{equation}\label{eq:partition-m}
Z_N(m) = \sum_{e,t=0}^{t_{\max}} P(m,e,N)(t) .
\end{equation}

Since $t$ is limited in our simulation (the simulation follows
the evolution of the spin configurations up to $t_{\max}$) the
stable states will be under represented in this summation. These
states will contribute to the infinitely deep well of the stable
equilibrium state.
Then the free energy for each value of magnetization is given by

\begin{equation}
F_N(m) = -(k_b T) \ln (Z_N(m)) 
\end{equation}

\noindent or in terms of the exchange coupling $J$

\begin{equation}
F_N(m) = -\frac{k_B T}{J}\ln (Z(m)) \qquad .
\end{equation}

\begin{figure}[ht]
  \begin{center}
{\epsfig{file=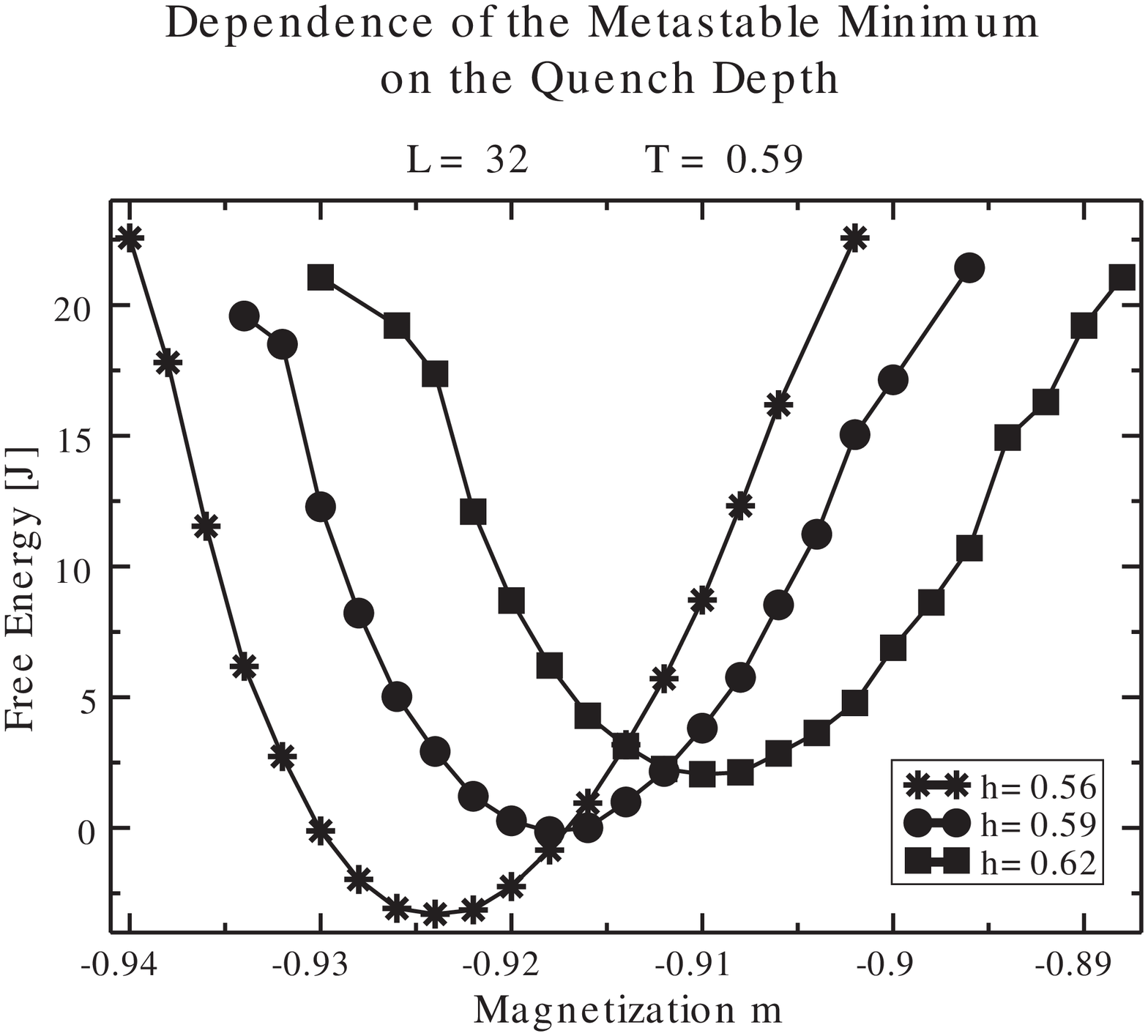,width=14cm}}
\caption{\label{fig:fe-min-32}Shown is the free energy minimum which results from
a constraint of the summation over time. Here the summation was constrained to
include only those times where the magnetization did not drop significantly
towards the equilibrium value. Shown are the results for the smallest system size 
($L=32$) and three magnetic fields h}
\end{center}
\end{figure}

If we  constrain the summation in Eq.~\ref{eq:partition-m} to
those times where the system is fluctuating around the metastable
magnetization, we would only obtain the metastable minimum of the
free energy as shown in Figure~\ref{fig:fe-min-32}.

\begin{figure}[ht]
  \begin{center}
{\epsfig{file=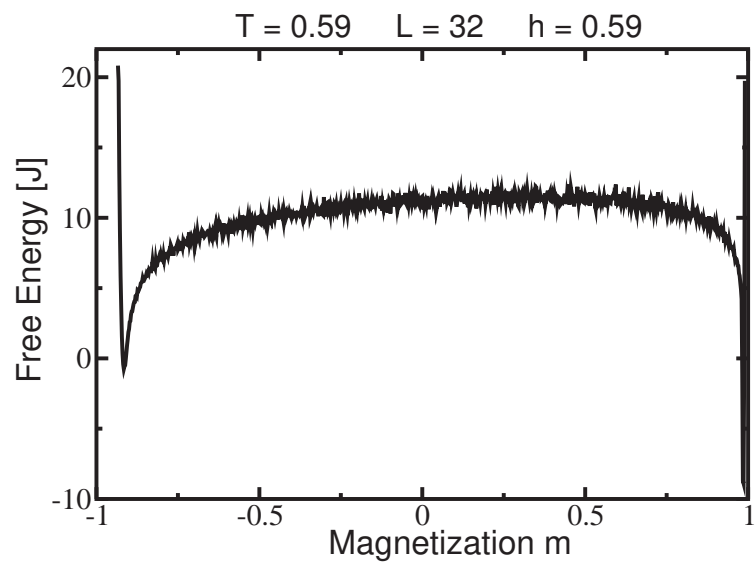,width=12cm}}
\caption{\label{fig:fe-one}Free energy for a particular quench depth
(strength of the magnetic field $h$). }
\end{center}
\end{figure}

An example of the full dependence of the free energy on $m$ is
shown in Figure~\ref{fig:fe-one} for one quench depth and
temperature. The fact that the minimum of the stable equilibrium state is
not infinitely deep is due to the limited summation as discussed
above. What is surprising is that the barrier extends over the
entire unstable region. Mean field theories suggest a different
picture as will be discussed in the next section.

\begin{figure}[ht]
  \begin{center}
{\epsfig{file=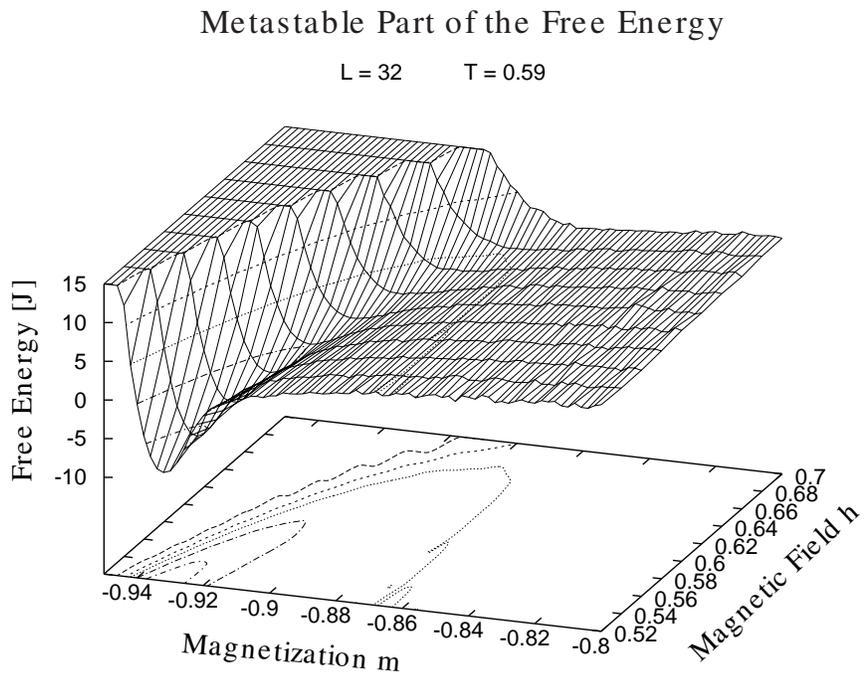,width=12cm}}
\caption{\label{fig:fe}Development of the free energy as a function of the quench depth
(strength of the magnetic field $h$)}
\end{center}
\end{figure}

The shift in the metastable minimum as a function of the quench
depth is shown in  Figure~\ref{fig:fe}. Note that the minimum of
the free energy widens considerably as we probe deeper into the
metastable/unstable region reflecting the fact that the paths
the system takes fluctuate much more than for small $h$. Also the
barrier height reduces with increasing $h$.

Most important for the consideration whether there is a clear cut
distinction between metastable and unstable states (that is, a
spinodal) we have calculated the barrier height in the free
energy. The barrier height is defined as the difference between
the minimum value of the free energy in the metastable state and
the maximum value of the free energy in the unstable part

\begin{equation}
\Delta F_N(m) = F_N(m_{max}) -  F_N(m_{min}) .
\end{equation}

\begin{figure}[ht]
  \begin{center}
{\epsfig{file=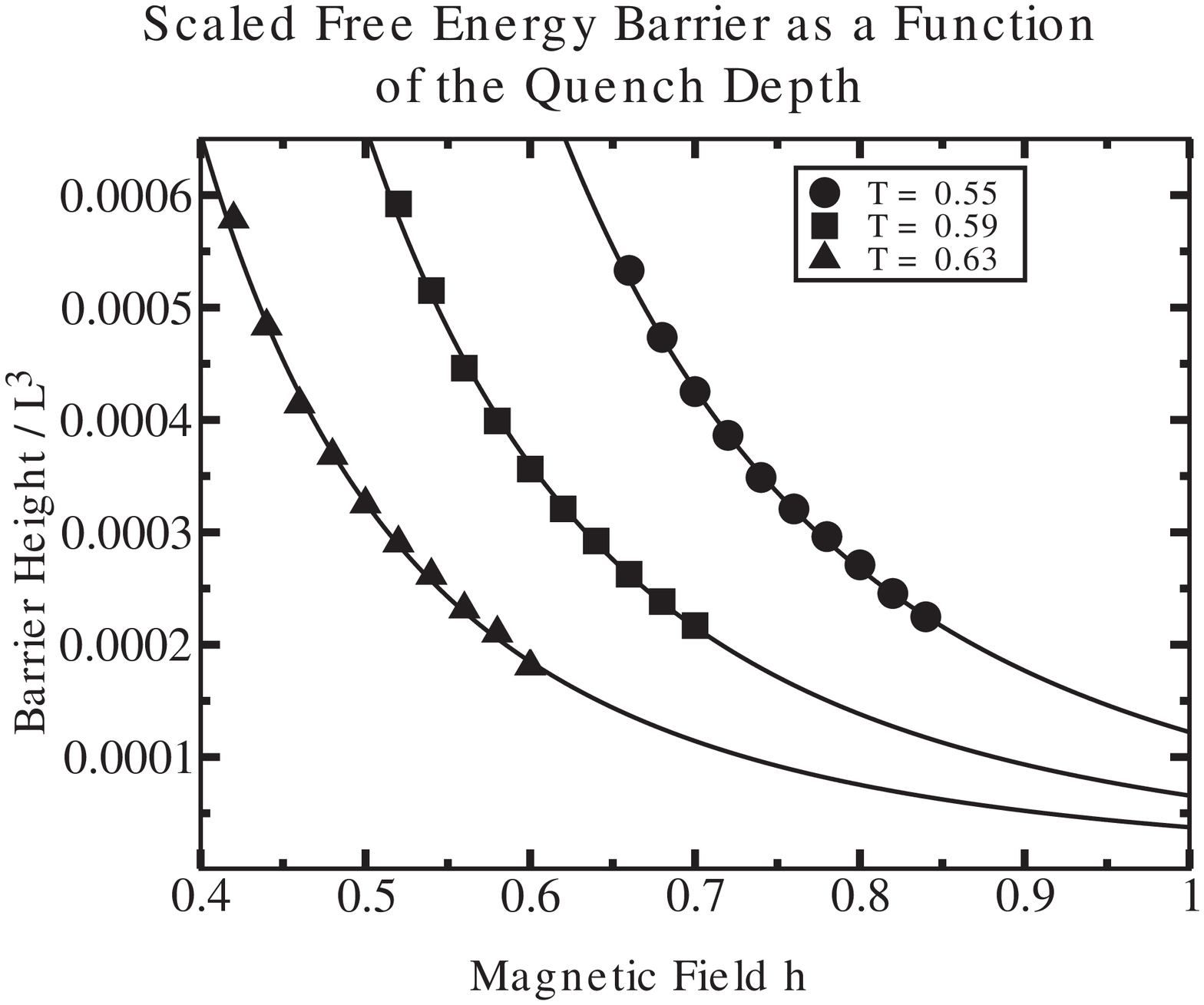,width=14cm}}
\caption{\label{fig:free-energy-barrier}Shown is the free energy barrier height as a function of the
quench depth (strength of the magnetic field $h$)}
\end{center}
\end{figure}

\noindent We divide out the volume dependence of the free energy and plot
in Figure~\ref{fig:free-energy-barrier} the free energy
barrier as a function of the temperature and the applied magnetic
field. The barrier height does not go to zero at a finite value of
the applied field as suggested by mean-field theories. Rather the
barrier height remains finite for those parameters where we still
can define a lifetime of a metastable state.

\begin{figure}[ht]
  \begin{center}
{\epsfig{file=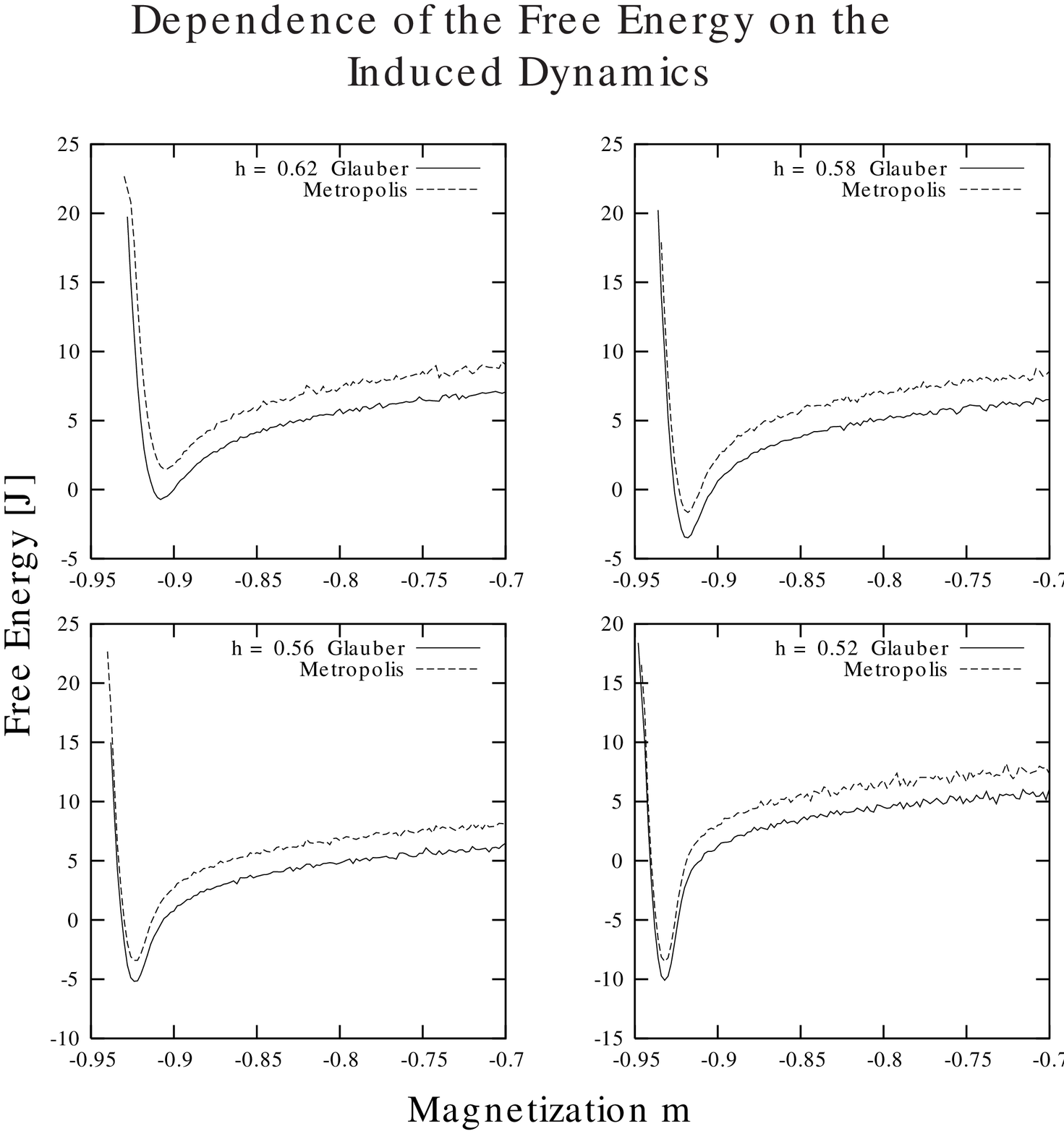,width=14cm}}
\caption{\label{fig:barrier-height}Shown is a comparison of the free energy as a 
function of the quench depth (strength of the magnetic field $h$) and the induced 
dynamics by the transition probabilities.}
\end{center}
\end{figure}

\begin{figure}[ht]
  \begin{center}
{\epsfig{file=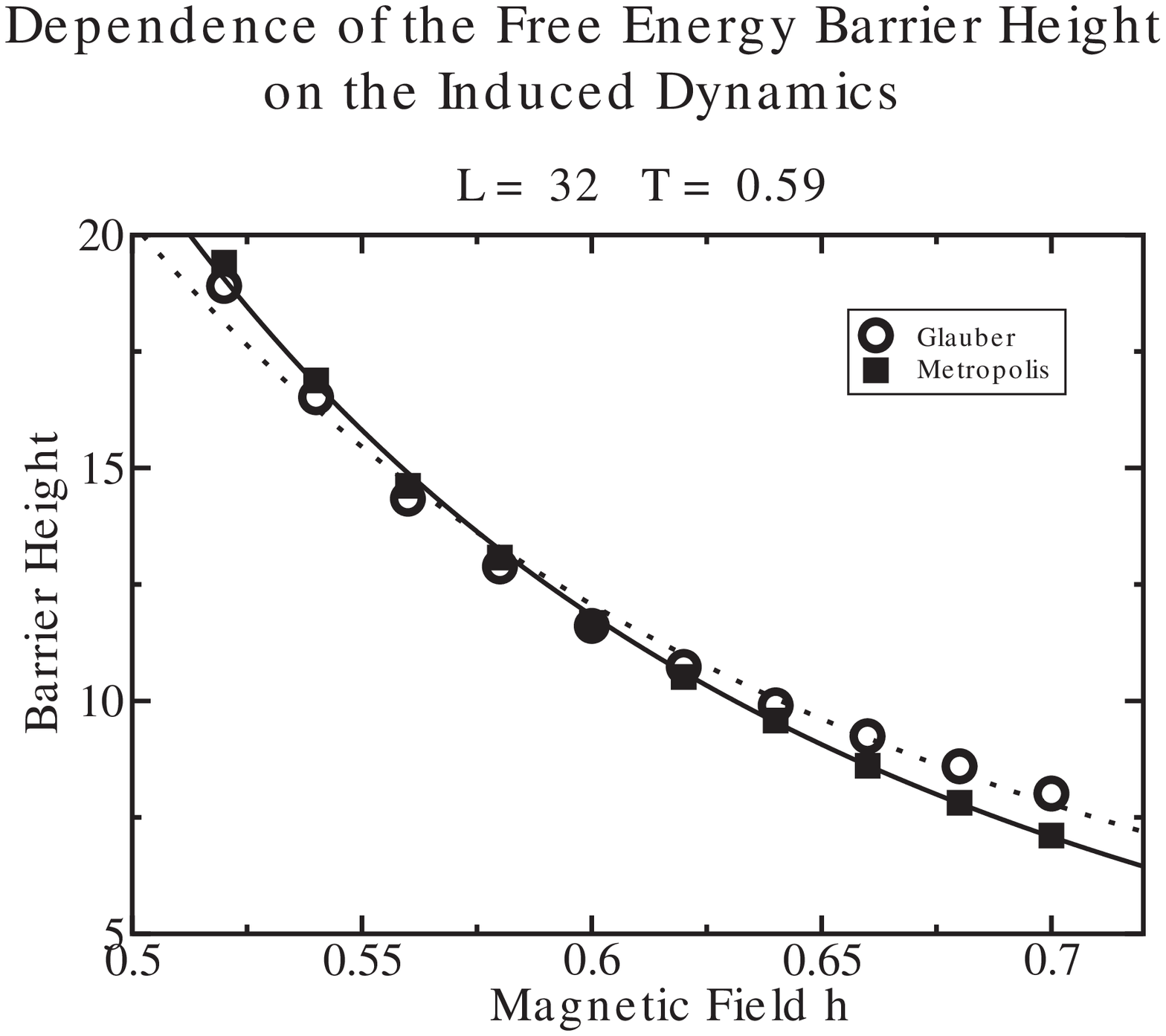,width=12cm}}
\caption{\label{fig:free-energy-both}Shown is a comparison of the free energy barrier 
height as a function of the quench depth (strength of the magnetic field $h$) and the
induced dynamics by the transition probabilities.}
\end{center}
\end{figure}

We now take a look at the dependence of the free energy on the
transition probabilities. In Figures~\ref{fig:barrier-height} and \ref{fig:free-energy-both} we
show the free energy for various magnetic fields for the two
transition probabilites. While the metastable minimum is not
effected by the choice of transition probability, we find the free
energy to be shifted in its absolute value. This is to be expected
because one can remove the metastable part of the phase diagram
altogether for example by using a cluster
algorithm~\cite{Burkitt-Heermann}. However, the dependence is not
very drastic (see below the discussion on the barrier height and
its relation to a mean-field spinodal). 

\section{Comparison to Mean-Field Theories}

In mean-field theory for the Ising model we derive the free energy
by considering the limit,  where all spins interact with each
other. In this situation the nearest-neighbour summation in the
Hamiltonian factorizes 

\begin{equation}
{\cal H}(m) = -\frac{1}{2}\frac{T_c}{T}m^2 -hm .
\end{equation}

\begin{figure}[ht]
  \begin{center}
{\epsfig{file=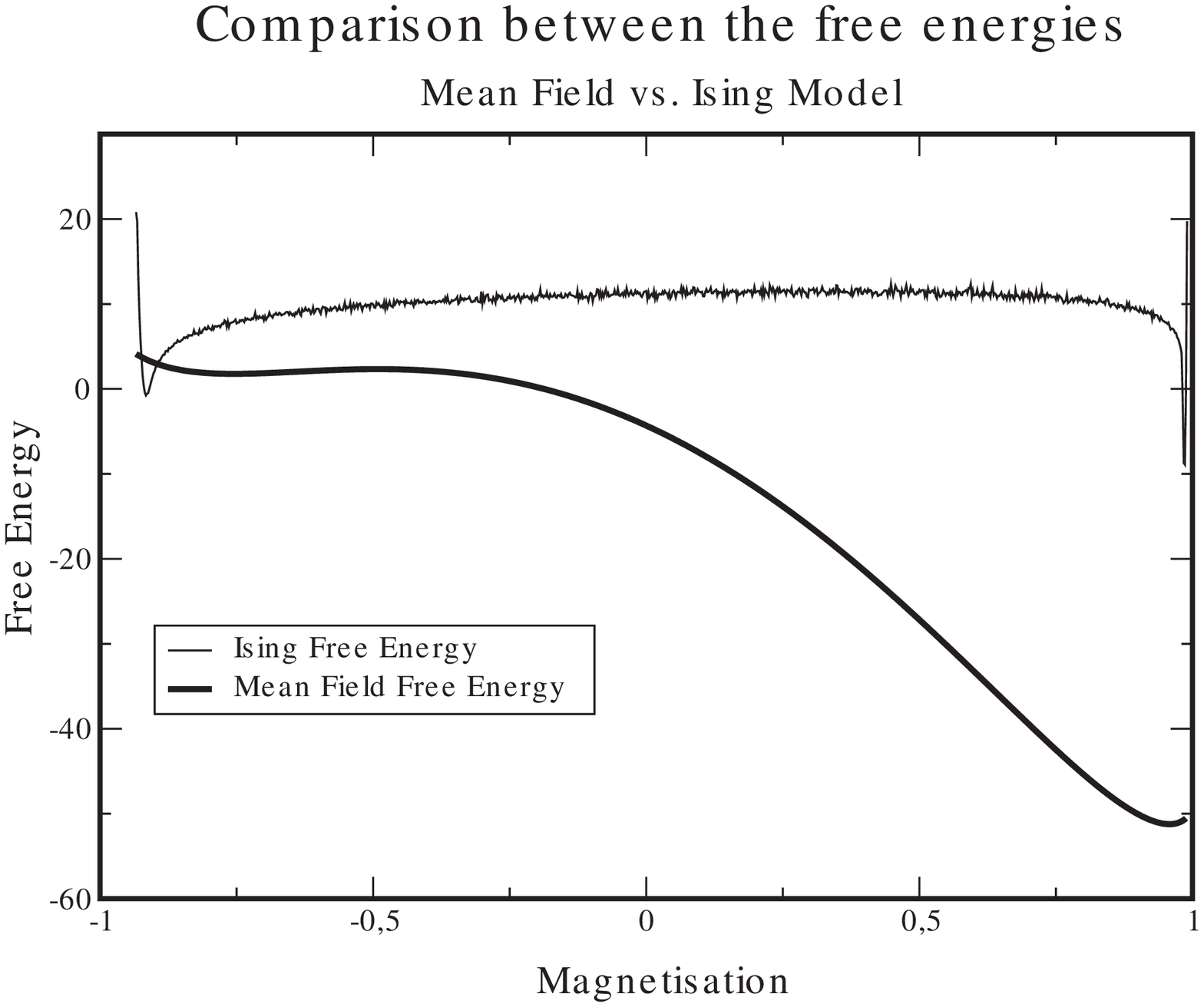,width=12cm}}
\caption{\label{fig:fe-mft}Comparison between the mean-field and the simulation result
of the free energy for a particular quench depth (strength of the magnetic field $h$).}
\end{center}
\end{figure}

\noindent We now consider the entropy of mixing of the states $m$ and $-m$
and find for the free energy the well known result

\begin{equation}
F(h,m) = (\frac{1+m}{2}\ln {\frac{1+m}{2}} +\frac{1-m}{2}\ln {\frac{1-m}{2}}) -\frac{1}{2}\frac{T_c}{T}m^2 -hm 
\end{equation}

\noindent for the free energy. A comparison between the predicted
mean-field theory and the result from the simulation (c.f.
Figure~\ref{fig:fe-mft}) shows clearly that not only 
the predicted metastable magnetization is wrong, but also the
form of the function between  the two minima. This holds also for
the predicted free energy barrier.

An important theoretical consequence from the above mean-field
free energy is the existence of a spinodal~\cite{Revs1,Revs2,Rikvold}. The
spinodal is defined to the loci of the points, where the free
energy barrier is zero. If there is a spinodal, then we would be
able to scale the results of the  barrier height as

\begin{equation}
\Delta F =4\left(  \frac{T}{T_c}\right) (h
-h_{\rm sp})^{x}|m_{\rm sp}|^{-1/2} .
\end{equation}

\noindent In mean-field theory the power of the exponent $x$ is 3/2.
However, no consistent set  of $h_{\rm sp}$, that is, assuming a
spinodal which not necessarily corresponds to the mean-field
spinodal, and exponent could be found. Hence, there is not 
a spinodal in the considered model. 

\section{Discussion}

We have shown that using the ensemble of relaxation paths one can
calculate the  free energy for finite range interaction models. 
The kind of analysis presented in this paper very strongly
suggests that there is no spinodal, where the free energy barrier
is zero, at least for models with short-range potential. We rather
find a gradual decrease of the height of the barrier well into the
believed unstable part of the two-phase region. It would be
interesting to see if one could reconstruct the free energy from
the quenches into the unstable region.

\section{Acknowledgment}
 
A.T.B and C.E.C would like to thank DAAD and FAPERJ for financial
support. We are also very grateful for discussions with D. Stauffer, 
H. Gould and K. Binder.

\vfill
\newpage

\vfill
\newpage

\vfill
\newpage

\vfill
\newpage

\vfill
\newpage

\vfill
\newpage

\vfill
\newpage

\vfill
\newpage

\vfill
\newpage

\vfill
\newpage

\end{document}